\documentclass[aps,prl,reprint,groupedaddress]{revtex4-1}
\usepackage{graphicx} 
\usepackage{dcolumn} 
\usepackage{bm}
\usepackage{hyperref}
\usepackage{amsmath}
\usepackage{amssymb}
\usepackage{float}

\begin{document}

\title{ Dynamics of light nuclei produced in the massive transfer reactions }
\author{Zi-Han Wang$^{1}$ }
\author{Zhao-Qing Feng$^{1,2}$ }
\email{Corresponding author: fengzhq@scut.edu.cn}
\affiliation{ $^{1}$School of Physics and Optoelectronics, South China University of Technology, Guangzhou 510640, China  \\
$^{2}$State Key Laboratory of Heavy Ion Science and Technology, Institute of Modern Physics, Chinese Academy of Sciences, Lanzhou 730000, China }

\date{\today}

\begin{abstract}
Within the framework of the dinuclear system (DNS) model by implementing the cluster transfer into the dissipation process, we systematically investigated the energy spectra and the angular distribution of the preequilibrium clusters (n, p, d, t, $^{3}$He, $\alpha$, $^{6,7}$Li, $^{8,9}$Be) in the massive transfer reactions of $^{12}$C+$^{209}$Bi, $^{14}$N+$^{159}$Tb, $^{14}$N+$^{169}$Tm, $^{14}$N+$^{181}$Ta, $^{14}$N+$^{197}$Au, $^{14}$N+$^{209}$Bi, $^{58,64,72}$Ni+$^{198}$Pt near the Coulomb barrier energies. It is found that the neutron emission is the most probable in comparison with the charged particles and the $\alpha$ yields are comparable with the hydrogen isotopes in magnitude. The preequilibrium clusters are mainly produced from the projectile-like and target-like fragments in the evolution of dinuclear system. The kinetic energy spectra manifest the Boltzmann distribution and the Coulomb potential influences the structure. The preequilibrium clusters follows the angular distribution of multinucleon transfer fragments.

\begin{description}
\item[PACS number(s)]
25.70.Hi, 25.70.Lm, 24.60.-k      
\end{description}
\end{abstract}

\maketitle

\section{I. Introduction}

The cluster structure in a atomic nucleus is a spatially located subsystem consisting of strongly related nucleons with much greater internal binding energy than external ones, which can be treated as a whole without considering its internal structure \cite{Wei1995}. In 1968, Ikeda proposed that the nuclear cluster states tend to occur in excited states near the cluster threshold energy \cite{Ik1968}. In some weakly bound nuclei, the cluster structure is more obvious, and the cluster structure is also ubiquitous in light nuclei, for example, the configuration of $^{6}$Li being composed of a $\alpha$ particle and a deuteron, the 2$\alpha$ structure for $^{8}$Be, 3$\alpha$ for $^{12}$C, 5 $\alpha$ for $^{20}$Ne \cite{Zh2023}. The most convenient way to study the cluster structure inside a nucleus is to separate the cluster via the pick-up or stripping reactions. The theoretical explanation of the preequilibrium reaction was initially developed through the exciton model. Semi-classical theories was less successful in explaining the angular distribution of emitted particles. The Boltzmann master equation theory was mainly used to calculate the energy spectra of particles emitted in nucleon-induced reactions and heavy ion reactions. Details can be found in the review Ref. \cite{Ho2003} and its references. On the other hand, the emission of preequilibrium clusters in transfer reactions around the Coulomb barrier is also an important physical problem. The emission of preequilibrium clusters is a complex process, which is related not only to the cluster structure of the collision system, but also to the dynamics of the reaction.In the treatment of nuclear structure, the cluster state is the overlap of the single particle wave functions. In a nuclear reaction, the formation of a preequilibrium particle is different from the cluster emitted during the de-excitation of a composite nucleus, and the pre-equilibrium cluster is formed before the formation of compound nucleus. Its emission continues until the formation of the composite nucleus, and the preequilibrium cluster may be emitted from any fragments during the reaction. Cluster emission provides important information for the study of single-particle states or multiparticle correlation of a nuclei, and it is a powerful tool for nuclear spectroscopy \cite{Ho2003}.

Since the multinucleon transfer (MNT) reactions and deep inelastic heavy ion collisions were proposed in the 1970s \cite{Ar1971, Bu1973, Ar1973}, a large number of experiments have been carried out to measure the double differential cross sections, the angular distributions and the energy distributions of different reaction systems. However, it is worth noting that there has been relatively little research on pre-equilibrium cluster emission in transfer reactions, both experimentally and theoretically. In the 1980s, scientists at RIKEN in Japan and at IMP in China have measured the preequilibrium cluster emission of  the transfer reactions of $^{14}$N+$^{159}$Tb, $^{169}$Tm, $^{181}$Ta, $^{197}$Au, $^{209}$Bi \cite{Ut1980} and $^{12}$C+$^{209}$Bi \cite{Sh1977,Jin1980}, respectively. The angular distributions, kinetic energy spectra and total production cross sections of the emitted particles were measured in experiments. As we all know, since the concept of superheavy stable island was proposed in the 1960s, the synthesis of superheavy nuclei has become an important frontier in the field of nuclear physics. In the past few decades, 15 kinds of the superheavy elements $Z=104\sim118$  \cite{Th2013} have been synthesized artificially by hot fusion or cold fusion reaction. However, due to the limitations of the projectile-target materials and experimental conditions, the fusion evaporation reaction is difficult to reach the next period of the periodic table. With the MNT reactions, we can generate many nuclei depending on the transfer channels, and the excitation energy of compound nuclei in MNT reactions distributes widely. With the development of the separation and detection technology, MNT reaction may be the most promising method to synthesize unknown superheavy elements. This mechanism has been applied to the production of heavy and superheavy isotopes \cite{De2022, Ni2023}.

The study of the pre-equilibrium cluster emission in the MNT reaction is not only of great significance for understanding the cluster structure of the collision system, but also for exploring the formation mechanism of the cluster, the kinetic information of the reaction process and the nuclear astrophysical process. The High Intensity Accelerator Facility (HIAF) built in Huizhou, China, has a large-energy range and a wide species of particle beams \cite{Ya13}, which provides a good experimental platform for the study of nuclear cluster structures and cluster emission.

In this work, we have systematically investigated the preequilibrium cluster emission in the transfer reactions. The article is organized as follows. In Section II we give a brief description of the DNS model for describing the preequilibrium cluster production. In Section III, the production cross sections, kinetic energy spectra and angular distributions of the preequilibrium clusters are analyzed and discussed. Summary and perspective are shown in Section IV.

\section{II. Brief description of the model}

The dinuclear system (DNS) model was first proposed by Volkov \cite{Vo1978} to describe the deep inelastic heavy ion collisions.
Adamian et al. applied the DNS model to fusion evaporation reactions in competition with the quasi-fission process to study the synthesis of superheavy nuclei for the first time \cite{Ad1996, Ad1997, Ad1998}. The Lanzhou nuclear physics group has further developed the DNS model \cite{Li2003, Feng2005, Feng2006, Feng2007}, e.g., introduced the barrier distribution function method in the capture process, considered the effects of quasi-fission and fission in the fusion stage, used statistical evaporation theory and Bohr-Wheeler formula to calculate the survival probability of superheavy nucleus. The DNS model has been used widely to study the production cross section, quasi-fission, fusion dynamics etc. in the synthesis of superheavy nucleus based on fusion evaporation (FE) reactions and the multinucleon transfer (MNT) reactions \cite{Feng2008, Feng2009, Feng2010}.

With the DNS model, we have calculated the temporal evolution, the kinetic energy spectra and the angular distributions of the pre-equilibrium clusters in transfer reactions with the incident energy near the Coulomb barrier. Compared to our previous work \cite{Feng2023}, we have introduced the transfer of clusters in the master equation of the DNS model, and the Coulomb force is considered in the pre-equilibrium cluster emission process.
The pre-equilibrium particle is formed before the compound nucleus is formed, and its emission continues until the composite nucleus is formed. The cross section of pre-equilibrium particle emission ($\nu=n, p, d, t$, $^{3}$He, $\alpha$, $^{6,7}$Li and $^{8,9}$Be) is defined as
\begin{eqnarray}
\sigma_{\nu}(E_{k},\theta,t)  && =  \sum^{J_{max}}_{J=0} \sum^{Z_{max}}_{Z_{1}=Z_{\nu}} \sum^{N_{max}}_{N_{1}=N_{\nu}} \pi \overline{\lambda}^{2} (2J+1) \int f(B)   \nonumber\\
&&  \times  T(E_{c.m.}, J, B) P(Z_{1},N_{1},E_{1}(E_{c.m.},J), t, B)                   \nonumber \\
&&  \times  P_{\nu}(Z_{\nu},N_{\nu},E_{k}) dB.
\end{eqnarray}
Here, the reduced de Broglie wavelength $\overline{\lambda}=\hbar/\sqrt{2\mu E_{c.m.}}$, and $P(Z_{1},N_{1},E_{1}(E_{c.m.},J), t, B)$ denotes the realization probability of the DNS fragment $(Z_{1},N_{1})$. $P_{\nu}(Z_{\nu},N_{\nu},E_{k})$ is the emission probability of the pre-equilibrium particles. $E_{1}$ is the excitation energy for the fragment $(Z_{1},N_{1})$, which is associated with the center-of-mass energy $E_{c.m.}$ and incident angular momentum $J$. The maximal angular momentum $J_{max}$ is taken to be the grazing collision of two colliding nuclei. The DNS fragments $(Z_{1},N_{1})$ ranging from the light one $(Z_{\nu},N_{\nu})$ to the composite system $(Z_{max},N_{max})$, with $Z_{max}=Z_{T}+Z_{P}$ and $N_{max}=N_{T}+N_{P}$ being the total proton and neutron numbers, respectively.

\subsection{2.1 The capture cross section of binary system overcoming the Coulomb barrier }
In the capture stage, the collision system overcomes the Coulomb barrier to form a composite system. The capture cross section is given by
\begin{eqnarray}
\sigma_{cap}(E_{c.m.}) & =  \pi \overline{\lambda}^{2} \sum^{J_{max}}_{J=0} (2J+1)      \nonumber \\
& \times \int f(B) T(E_{c.m.},J,B) dB,
\end{eqnarray}
where $T(E_{c.m.},J,B)$ is the penetration probability overcoming the barrier $B$. For the light and medium systems, $T(E_{c.m.},J)$ is calculated by the well-known Hill-Wheeler formula \cite{Hi1953},
\begin{eqnarray}
&& T(E_{\rm c.m.},J)= \nonumber \int   f(B) \\&& \frac{1}{1+\exp\left \{ -\frac{2\pi }{\hbar \omega (J)}\left [ E_{\rm c.m.}-B-\frac{\hbar^2J(J+1)}{2\mu R\rm_B^2(J) }  \right ]  \right \} } dB,
\end{eqnarray}
with $\hbar \omega (J)$ being the width of the parabolic barrier at $R_{B} (J)$. For the heavy systems, the collision system does not form a potential energy pocket after overcoming the Coulomb barrier, $T(E_{c.m.},J)$ is calculated by the classic trajectory method,
\begin{eqnarray}
T(E_{c.m.},J) = \begin{cases}
0,  \quad  E_{\text{c.m.}} < B + J(J+1)\hbar^2/(2\mu R^2_C),    \\
1,  \quad  E_{\text{c.m.}} > B + J(J+1)\hbar^2/(2\mu R^2_C).
\end{cases}
\end{eqnarray}
The reduced mass is $\mu = m_{n} A_{P} A_{T} /(A_{P} + A_{T})$ with $m_{n}$, $A_{P}$ and $A_{T}$ being the nucleon mass and the mass numbers of projectile and target nuclei, respectively. $R_{C}$ denotes the Coulomb radius.

The barrier distribution function is Gaussian form \cite{Feng2006, Za2001}
\begin{eqnarray}
f(B)= \frac{1}{N} exp[-((B-B_{m})/\Delta)^{2}].
\end{eqnarray}
The normalization constant satisfies $\int f(B)dB=1$. The quantities $B_{m}$ and $\Delta$ are evaluated by $B_{m}=(B_{C} + B_{S})/2$ and $\Delta = (B_{C} - B_{S})/2$, respectively. The $B_{C}$ is the Coulomb barrier at waist-to-waist orientation and $B_{S}$ is the minimum barrier by varying the quadrupole deformation of the colliding partners. Here we take $B_{S}$ as the Coulomb barrier at tip-to-tip orientation.

\subsection{2.2 The nucleon and cluster transfer dynamics}
In the nucleon transfer process, the distribution probability of the DNS fragments is obtained by numerically solving a set of master equations \cite{Fe2023}. Fragment ($Z_{1}$,$N_{1}$) has the proton number of $Z_{1}$, the neutron number of $N_{1}$, the internal excitation energy of $E_{1}$, and the quadrupole deformation $\beta_{1}$, and the time evolution equation of its distribution probability can be described as
\begin{widetext}
\begin{eqnarray}
&&  \frac{d P(Z_{1},N_{1},E_{1},\beta_{1},B,t)}{dt}     \nonumber\\
&&  = \sum_{Z^{'}_{1} = Z_{1} \pm 1}W^{p}_{Z_{1},N_{1},\beta_{1};Z^{'}_{1},N_{1},\beta^{'}_{1}}(t) \times [d_{Z_{1},N_{1}}P(Z^{'}_{1},N_{1},E^{'}_{1},\beta^{'}_{1},B,t) - d_{Z^{'}_{1},N_{1}}P(Z_{1},N_{1},E_{1},\beta_{1},B,t)] \nonumber\\
&&  + \sum_{N^{'}_{1} = N_{1} \pm 1}W^{n}_{Z_{1},N_{1},\beta_{1};Z_{1},N^{'}_{1},\beta^{'}_{1}}(t) \times [d_{Z_{1},N_{1}}P(Z_{1},N^{'}_{1},E^{'}_{1},\beta^{'}_{1},B,t) - d_{Z_{1},N^{'}_{1}}P(Z_{1},N_{1},E_{1},\beta_{1},B,t)] \nonumber\\
&&  + \sum_{Z^{'}_{1} = Z_{1} \pm 1, N^{'}_{1} = N_{1} \pm 1}W^{d}_{Z_{1},N_{1},\beta_{1};Z_{1},N^{'}_{1},\beta^{'}_{1}}(t) \times [d_{Z_{1},N_{1}}P(Z^{'}_{1},N^{'}_{1},E^{'}_{1},\beta^{'}_{1},B,t) - d_{Z^{'}_{1},N^{'}_{1}}P(Z_{1},N_{1},E_{1},\beta_{1},B,t)] \nonumber\\
&&  + \sum_{Z^{'}_{1} = Z_{1} \pm 1, N^{'}_{1} = N_{1} \pm 2}W^{t}_{Z_{1},N_{1},\beta_{1};Z_{1},N^{'}_{1},\beta^{'}_{1}}(t) \times [d_{Z_{1},N_{1}}P(Z^{'}_{1},N^{'}_{1},E^{'}_{1},\beta^{'}_{1},B,t) - d_{Z^{'}_{1},N^{'}_{1}}P(Z_{1},N_{1},E_{1},\beta_{1},B,t)] \nonumber\\
&&  + \sum_{Z^{'}_{1} = Z_{1} \pm 2, N^{'}_{1} = N_{1} \pm 1}W^{^{3}He}_{Z_{1},N_{1},\beta_{1};Z_{1},N^{'}_{1},\beta^{'}_{1}}(t)  \times [d_{Z_{1},N_{1}}P(Z^{'}_{1},N^{'}_{1},E^{'}_{1},\beta^{'}_{1},B,t) - d_{Z^{'}_{1},N^{'}_{1}}P(Z_{1},N_{1},E_{1},\beta_{1},B,t)] \nonumber\\
&&  + \sum_{Z^{'}_{1} = Z_{1} \pm 2, N^{'}_{1} = N_{1} \pm 2}W^{\alpha}_{Z_{1},N_{1},\beta_{1};Z_{1},N^{'}_{1},\beta^{'}_{1}}(t) \times [d_{Z_{1},N_{1}}P(Z^{'}_{1},N^{'}_{1},E^{'}_{1},\beta^{'}_{1},B,t) - d_{Z^{'}_{1},N^{'}_{1}}P(Z_{1},N_{1},E_{1},\beta_{1},B,t)].
\end{eqnarray}
\end{widetext}

And in the equation, $W_{Z_{1},N_{1},\beta_{1};Z_{1}^{\prime},N_{1}^{\prime},\beta_{1}^{\prime}}$ is the mean transition probability from the channel $(Z_{1},N_{1},E_{1},\beta_{1})$ to $(Z_{1}^{\prime},N_{1}^{\prime},E_{1}^{\prime},\beta_{1}^{\prime})$.
The quantity $d_{Z_{1},N_{1}}$ indicates the microscopic dimension corresponding to the macroscopic state $(Z_{1},N_{1},E_{1},\beta_{1})$.
In this process, the transfer of nucleons or clusters is satisfied with the relationships, $Z_{1}^{\prime} = Z_{1} \pm Z_{\nu}$ and $N_{1}^{\prime} = N_{1} \pm N_{\nu}$, each represents the transfer of a neutron, proton, deuteron, tritium, $^{3}$He, $\alpha$.
Note that we ignored the quasi-fission of DNS and the fission of heavy fragments in the dissipation process.
The initial probabilities of projectile and target nuclei are set to be $P(Z_{proj},N_{proj},E_{1}=0, t=0) = P(Z_{targ},N_{targ},E_{1}=0, t=0)=0.5$.
The nucleon transfer process satisfies the unitary condition $\sum_{Z_{1},N_{1}} P(Z_{1}, N_{1}, E_{1}, t)=1$.

Similar to the cascade transfer of nucleons \cite{Feng2006}, the transfer of clusters is also described by the single-particle Hamiltonian,
\begin{eqnarray}
H(t)= H_{0}(t) + V(t).
\end{eqnarray}
Single-particle states are defined with respect to the centers of the interacting nuclei and are assumed to be orthogonalized in the overlap region. Thus, the annihilation and creation operators depend on the time.
The total single particle energy is
\begin{eqnarray}
H_0(t) &&= \sum _K\sum_{\nu_K} \varepsilon_{\nu_K}(t)\alpha^+_{\nu_K}(t)\alpha_{\nu_K}(t).
\end{eqnarray}
The interaction potential is
\begin{eqnarray}
V(t) &&= \sum _{K,K^{'}} \sum_{\alpha_K,\beta_{K'}} u_{\alpha_K,\beta_{K'}}\alpha^+_{\alpha_K}(t)\alpha_{\beta_K}(t) \nonumber  \\
&&= \sum_{K,K'}V_{K,K'}(t).
\end{eqnarray}
The quantity $\varepsilon_{\nu K}$ represents the single-particle energies, and $u_{\alpha_K,\beta_{K'}}$ is the interaction matrix elements, parameterized to the following form:
\begin{eqnarray}
&& u_{\alpha_K,\beta_K'} =  U_{K,K'}(t)  \\ && \times \left\{ \exp \left[- \frac{1}{2}( \frac{\varepsilon_{\alpha_K}(t) - \varepsilon_{\beta_K}(t)}{\Delta_{K,K'}(t)})^2 \right] - \delta_{\alpha_K,\beta_{K'}} \right\}, \nonumber
\end{eqnarray}
here, the calculation of the $U_{\rm K,K'}(t)$ and $\delta_{\alpha_{\rm K},\beta_{\rm K'}}(t)$ have been described in Ref. \cite{Ad2003}.

In the relaxation process of the relative motion, the DNS will be excited by the dissipation of the relative kinetic energy and angular momentum. The excited DNS opens a valence space in which the valence nucleons have a symmetrical distribution around the Fermi surface. Only the particles at the states within the valence space are actively excited and undergo transfer. The averages on these quantities are performed in the valence space as follows:
\begin{eqnarray}
\Delta \varepsilon_K = \sqrt{\frac{4\varepsilon^*_K}{g_K}},\quad
\varepsilon^*_K =\varepsilon^*\frac{A_K}{A}, \quad
g_K = A_K /12,
\end{eqnarray}
here, the symbol $\varepsilon^{*}$ is the local excitation energy of the DNS fragments, which provides the excitation energy for the mean transition probability. The number of valence states in the valence space is $N_K$ = $g_{\rm K}\Delta\varepsilon_{\rm K}$, $g_{\rm K}$ is the single particle level density around the Fermi surface. The number of valence nucleon is $m_{\rm K}$ = $N_{\rm K/2}$.
The microscopic dimension for the fragment ($Z_{K},N_{K}$) is evaluated by
\begin{eqnarray}
 d(m_1, m_2) = {N_1 \choose m_1} {N_2 \choose m_2}.
\end{eqnarray}

The mean transition probability is related with the local excitation energy and the transfer of nucleons or clusters, and it can be microscopically derived from the interaction potential in valence space as
\begin{eqnarray}
&& W^{\nu}_{Z_{1},N_{1};Z_{1}^{\prime},N^{\prime}_{1}} = G_{\nu} \frac{\tau_{mem}(Z_{1},N_{1},E_{1};Z_{1}^{\prime},N_{1}^{\prime},E_{1}^{\prime})}{d_{Z_{1},N_{1}} d_{Z_{1}^{\prime},N_{1}^{\prime}}\hbar^{2}}    \nonumber \\
&& \times \sum_{ii^{\prime}}|\langle  Z_{1}^{\prime},N_{1}^{\prime},E_{1}^{\prime},i^{\prime}|V|Z_{1},N_{1},E_{1},i \rangle|^{2}.
\end{eqnarray}
$G_{\nu}$ represents the spin-isospin statistical factors, and we use the winger density approach to identify particle types \cite{Ma1997, Feng2020}, i.e. $G_{\nu} = 1, 1, 3/8, 1/12, 1/12, 1/96$ for neutron, proton, deuteron, tritium, $^{3}$He, $\alpha$, respectively.

The memory time is connected with the internal excitation energy \cite{Ri1979},
\begin{eqnarray}
&& \tau_{mem}(Z_1,N_1,E_1; Z'_1,N_1, E'_1) =   \nonumber \\
&& \left[\frac{2\pi \hbar^2} {\sum _{KK'} <V_{KK} V^*_{KK'}>}\right]^{1/2},
\end{eqnarray}
\begin{eqnarray}
&& <V_{KK} V^*_{KK'}> = \frac{1}{4} U^2_{KK'}g_K g_K' \Delta_{KK'} \Delta \varepsilon_K \Delta \varepsilon_K^{\prime}
 \nonumber \\
&&  \times \left[ \Delta^2_{KK'}+ \frac{1}{6} ((\Delta \varepsilon_K)^2  + (\Delta \varepsilon_K^{\prime})^2) \right]^{-1/2}.
\end{eqnarray}

The interaction matrix elements are calculated by
\begin{eqnarray}
\sum _{ii'} |V_{ii'}|^2 &&  = \omega_{11}(i_1,i'_1)  + \omega_{22}(i_1,i'_1)   \nonumber \\
&& + \omega_{12}(i_1,i'_1) + \omega_{21}(i_1,i'_1),
\end{eqnarray}
in which
\begin{eqnarray}
\omega_{KK'} (i,i'_1)=d_{Z_1,N_1} <V_{KK'},V^*_{KK'}>,
\end{eqnarray}
with the states $i(Z_1,N_1,E_1)$ and $i^{'}(Z^{'}_1,N^{'}_1,E^{'}_1)$.

In the relaxation process of the relative motion, the DNS will be excited by the dissipation of the relative kinetic energy. The local excitation energy is determined by the dissipation energy from the relative motion and the potential energy surface of the DNS \cite{Feng2007, Feng2009},
\begin{eqnarray}
\varepsilon^{\ast}(t)=E_{diss}(t)-\left(U(\{\alpha\})-U(\{\alpha_{EN}\})\right).
\end{eqnarray}
where ${\alpha_{EN}}={Z_{P},N_{P},Z_{T},N_{T},J,R,\beta_{P},\beta_{T},\theta_{P},\theta_{T}}$ for the projectile-target system.
The excitation energy of the DNS fragment$(Z_{1},N_{1})$ is $E_{1}=\varepsilon^{\ast}(t=\tau_{int})A_{1}/A$.
$\tau_{int}$ denotes the interaction time, which is associated with the reaction system and the relative angular momentum, and can be gained by the deflection function \cite{Wo1978}.
The energy dissipated into the DNS is
\begin{equation}
E_{diss}(t)=E_{c.m.}-B-\frac{\langle  J(t)\rangle(\langle J(t)\rangle+1)\hbar^{2}}{2\zeta_{rel}}-\langle  E_{rad}(J,t)\rangle.
\end{equation}
And the radial energy is
\begin{equation}
\langle  E_{rad}(J,t)\rangle=E_{rad}(J,0)\exp(-t/\tau_{r}),
\end{equation}
with the relaxation time of radial motion being $\tau_{r} = 5 \times 10^{-22}$ s, the initial radial energy being $E_{rad}(J,0)=E_{c.m.}-B-J_{i}(J_{i}+1)\hbar^{2}/(2\zeta_{rel})$.
The dissipation of the relative angular momentum is described by
\begin{equation}
\langle  J(t)\rangle=J_{st}+(J_{i}-J_{st})\exp(-t/\tau_{J})
\end{equation}
The angular momentum at the sticking limit is $J_{st}=J_{i}\zeta_{rel}/\zeta_{tot}$ and the relaxation time $\tau_{J}=15\times10^{-22}$ s.
The $\zeta_{rel}$ and $\zeta_{tot}$ are the relative and total moments of inertia of the DNS, respectively.
The initial angular momentum is set to be $J_{i}=J$ in Eq. (1).
The relaxation time of radial kinetic energy and angular momentum is associated with the friction coefficients in the binary collisions. The values in this work are taken from the empirical analysis in deeply inelastic heavy-ion collisions \cite{Wo1978, Li1983}.

The potential energy surface (PES) of the DNS is evaluated as
\begin{eqnarray}
U(\{\alpha\}) && = B(Z_{1},N_{1}) + B(Z_{2},N_{2})   \nonumber \\
&& - B(Z,N) + V(\{\alpha\}),
\end{eqnarray}
with the relationship of $Z_{1} + Z_{1} = Z$ and $N_{1} + N_{1} = N$ \cite{Fe2009, Feng2017}. The symbol $\{\alpha\}$ denotes the quantities ${Z_{1},N_{1},Z_{2},N_{2},J,R,\beta_{1},\beta_{2},\theta_{1},\theta_{2}}$. In the calculation, the distance $R$ between the centers of the two fragments is chosen to be the value at the touching configuration, in which the DNS is assumed to be formed.
The $B(Z_{i},N_{i}) (i=1,2)$ and $B(Z,N)$ are the negative binding energies of the fragment $(Z_{i},N_{i})$ and the compound nucleus $(Z,N)$, respectively.
The $\beta_{i}$ represent the quadrupole deformations of the two fragments at ground state, and $\theta_{i} (i=1,2)$ denote the angles between the collision orientations and the symmetry axes of deformed nuclei.
The interaction potential between fragment $(Z_{1},N_{1})$ and $(Z_{2},N_{2})$ is derived from
\begin{eqnarray}
V(\{\alpha\}) = V_{C}(\{\alpha\}) + V_{N}(\{\alpha\}) + V_{def}(t),
\end{eqnarray}
where $V_{C}$ is the Coulomb potential using the Wong formula \cite{Wo1973}, $V_{N}$ is the nucleus-nucleus potential using the double folding potential \cite{Go2017}, and $V_{def}(t)$ denotes the deformation energy of the DNS at reaction time t,
\begin{eqnarray}
V_{def}(t) = \frac{1}{2} C_{1} (\beta_{1} - \beta^{'}_{T}(t))^{2} + \frac{1}{2} C_{2} (\beta_{2} - \beta^{'}_{P}(t))^{2} .
\end{eqnarray}
The quantity $C_{i} (i=1,2)$ denote the stiffness parameters of the nuclear surface, which are calculated by the liquid drop model \cite{My1966}. Detailed calculations of $V_{def}(t)$ can be obtained from Ref. \cite{Chen2023} and the references therein.
Shown in Fig. 1 is the PES in collisions of $^{14}$N+$^{209}$Bi and $^{64}$Ni+$^{198}$Pt. The zigzag lines are the driving potentials, which are estimated by the minimal PES values in the process of transferring nucleons. The incident point is denoted by the star symbol.

\begin{figure*}
\includegraphics[width=\textwidth]{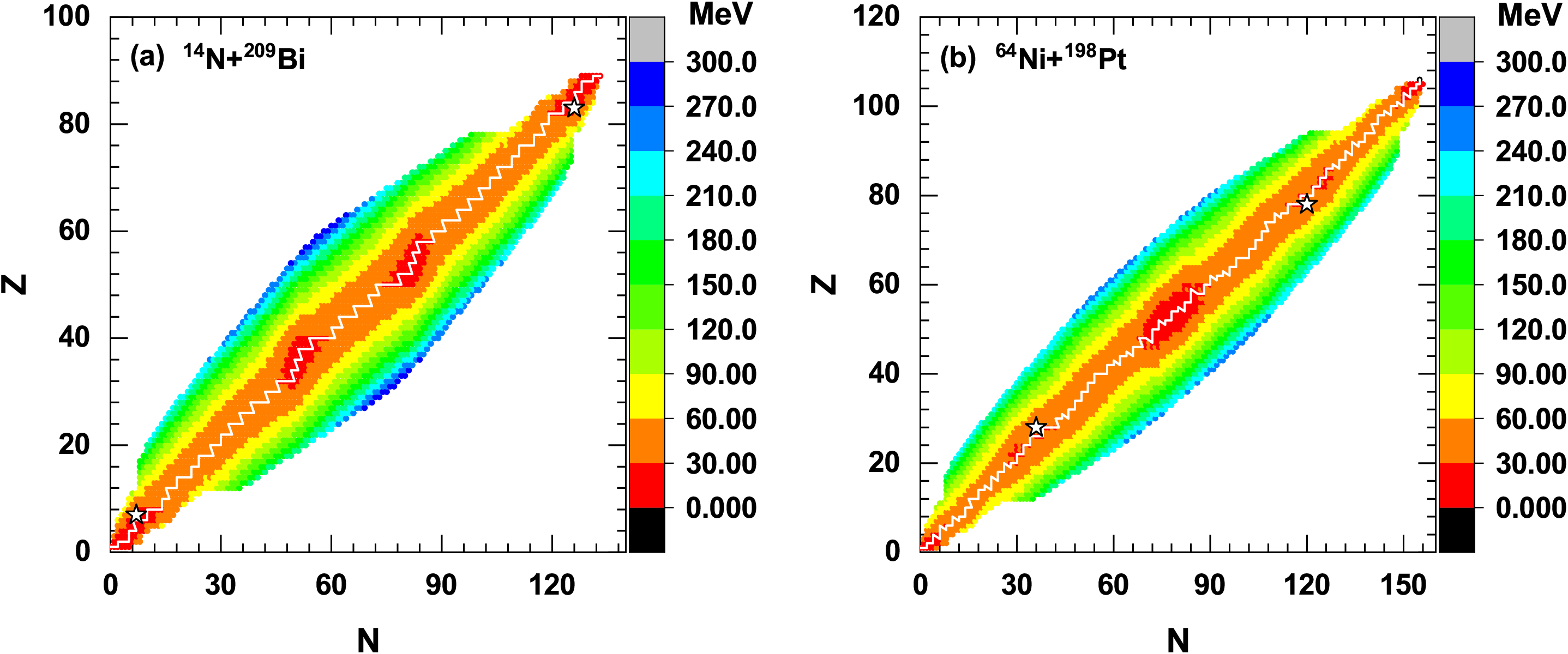}
\caption{The potential energy surfaces in the reactions of (a) $^{14}$N+$^{209}$Bi and (b) $^{64}$Ni+$^{198}$Pt, respectively. }
\end{figure*}

\subsection{2.3 The preequilibrium cluster emission }
The emission probabilities of pre-equilibrium clusters with the kinetic energy $E_{k}$ are calculated by the uncertainty principle within the time step $t \sim  t + \triangle t$ via
\begin{equation}
P_{\nu}(Z_{\nu},N_{\nu},E_{k}) = \triangle t\Gamma_{\nu}/\hbar. \label{emission Eq}
\end{equation}
Here the time step is set to be $\triangle t=0.5\times 10^{-22}$s for the reactions induced by $^{12}$C and $^{14}$N, but $\triangle t=0.25\times 10^{-22}$s for the reactions induced by $^{58,64,72}$Ni isotopes.

Based on the Weisskopf evaporation theory \cite{We1937, Chen2016}, we have the particle decay widths as follows,
\begin{eqnarray}
&& \Gamma_\nu(E^*,J) = (2s_\nu + 1) \frac{m_\nu}{\pi^2 \hbar^2 \rho(E^*,J)}\int \limits ^{E^* - B_\nu - E_{rot} - V_{c}}_0        \nonumber   \\
&&  \times  \varepsilon \rho(E^*-B_\nu - E_{rot} - V_{c} - \varepsilon, J)\sigma_{inv}(\varepsilon) d \varepsilon,
\end{eqnarray}
with $s_\nu$, $m_\nu$ and $B_\nu$ being the spin, mass and binding energy of the evaporating particles, respectively.

The inverse cross section is given by
\begin{eqnarray}
\sigma_{inv} = \pi R_\nu^{2}T(\nu),
\end{eqnarray}
with the radius of
\begin{eqnarray}
R_\nu = 1.21\left[(A-A_{\nu})^{1/3} + A_{\nu}^{1/3}\right].
\end{eqnarray}
The penetration probability is set to be $T(\nu) = 1$ for neutron and $T(\nu) = [1 + \exp(2\pi(V_C(\nu) - \varepsilon)/\hbar\omega)]^{-1}$ for charged particles with $\hbar \omega= 5$ MeV and 8 MeV for hydrogen isotopes and other charged particles, respectively.
It should be mentioned that the local equilibrium of the DNS is assumed to be formed and the excitation energy $E^*_{i} = \varepsilon^*_i $ for the $i-$th fragment is associated with the local excitation energy with the mass table \cite{Mo1995}.

The level density is calculated from the Fermi-gas model as
\begin{eqnarray}
\rho(E^{\ast},J) && = \frac{2J+1}{24\sqrt{2}\sigma^3a^{1/4}(E^{\ast} - \delta)^{5/4}}   \nonumber \\
&& \times\exp\left[ 2\sqrt{a(E^{\ast}-\delta)} - \frac{(J+1/2)^2}{2\sigma^2}\right],
\end{eqnarray}
with $\sigma^2 = 6\bar{m}^2\sqrt{a(E^*-\delta)}/\pi^2$ and $\bar{m}\approx0.24A^{2/3}$.
The pairing correction energy $\delta$ is set to be $12/\sqrt{A}, 0$ and $-12/\sqrt{A}$ for even-even, even-odd and odd-odd nuclei, respectively.
The level density parameter is related to the shell correction energy $E_{sh}(Z,N)$ and the excitation energy $E^{\ast}$ of the nucleus as
\begin{equation}
a(E^{\ast},Z,N)=\tilde{a}(A)[1+E_{sh}(Z,N)f(E^{\ast}-\Delta)/(E^{\ast}-\Delta)].
\end{equation}
The asymptotic Fermi-gas value of the level density parameter at high excitation energy is $\tilde{a}(A) = \alpha A + \beta  A^{2/3}b_{s}$, and the shell damping factor is given by $f(E^{\ast}) = 1-\exp(-\gamma E^{\ast})$ with $\gamma=\tilde{a}/(\epsilon  A^{4/3})$.
The parameters $\alpha$, $\beta$, $b_{s}$ and $\epsilon$ are taken to be 0.114, 0.098, 1. and 0.45, respectively \cite{Fe2009, Feng2017}.

The kinetic energy of the pre-equilibrium particle is sampled by the Monte Carlo method within the energy range $\epsilon_{\nu}\in(0, E^* - B_\nu -V_{C} - E_{rot})$. Here, $V_{C}$ represents the Coulomb force that the outgoing particles need to overcome, and for neutron, $V_{C}=0$.
Watt spectrum is used for the neutron emission \cite{Ro1992} and expressed as
\begin{equation}
\frac{dN_{n}}{d\epsilon_{n}}=C_{n}\frac{\epsilon_{n}^{1/2}}{T_{w}^{3/2}}\exp\left(-\frac{\epsilon_{n}}{T_{w}}\right)
\end{equation}
with $T_{w}=1.7\pm0.1$ MeV and normalization constant $C_{n}$.
For the charged particles, the Boltzmann distribution is taken into account as
\begin{equation}
\frac{dN_{\nu}}{d\epsilon_{\nu}}=8\pi E_{k} \left(\frac{m}{2\pi T_{\nu}}\right)^{1/2} \exp\left(-\frac{\epsilon_{\nu}}{T_{\nu}}\right),
\end{equation}
with $T_{\nu}=\sqrt{E^{\ast}/a}$, and $a=A/8$ being the level density parameter.

We use the deflection function method \cite{Wo1978, Peng2022} to calculate the angular distribution of the pre-equilibrium particles emitted from the DNS fragments as
\begin{eqnarray}
\Theta(J_{i}) = \Theta_{C}(J_{i}) + \Theta_{N}(J_{i}).
\end{eqnarray}
The Coulomb deflection is given by the Rutherford function as
\begin{eqnarray}
\Theta(J_{i})_{C}=2\arctan \frac{Z_{p}Z_{t}e^{2}}{2E_{c.m.} b}
\end{eqnarray}
with the incident energy $E_{c.m.}$ and impact parameter $b$.
The nuclear deflection is calculated by
\begin{eqnarray}
\Theta(J_{i})_{N} = -\beta\Theta_{C}^{gr}(J_{i}) \frac{J_{i}}{J_{gr}}\left(\frac{\delta}{\beta}\right)^{J_{i}/J_{gr}}.
\end{eqnarray}
Here $\Theta_{C}^{gr}(J_{i})$ is the Coulomb scattering angle at the grazing angular momentum $J_{gr}=0.22R_{int}[A_{red}(E_{c.m.}-V(R_{int}))]^{1/2}$.
The quantity $J_{i}$ is the incident angular momentum, $A_{red}$ is the reduced mass of the collision system, and $V(R_{int})$ denotes the interaction potential with $R_{int}$ being the Coulomb radius.
The parameters $\delta$ and $\beta$ are parameterized by fitting the deep inelastic scattering in massive collisions as
\begin{eqnarray}
\beta  = \begin{cases}
75 f(\eta) + 15 & \eta < 375, \\
36 \exp(-2.17\times 10 ^{-3} \eta) & \eta \geq 375.
\end{cases}
\end{eqnarray}
and
\begin{eqnarray}
\delta  = \begin{cases}
0.07 f(\eta) + 0.11  & \eta < 375, \\
0.117 \exp(-1.34\times 10 ^{-4} \eta) & \eta \geq 375.
\end{cases}
\end{eqnarray}
with
\begin{equation}
f(\eta) = [1 + \exp{\frac{\eta-235}{32}}]^{-1}.
\end{equation}
The quantity $\eta = \frac{Z_1 Z_2 e^2}{\upsilon}$, is the Sommerfeld parameter, and the relative velocity is calculated by $\upsilon = {\frac{2}{A_{red}}(E_c.m. - V(R_{int}))}^{1/2}$.
For the $i-$th DNS fragment, the emission angle is determined by
\begin{equation}
\Theta_{i}(J_{i}) = \Theta(J_{i}) \frac{\xi_{i}}{(\xi_{1}+\xi_{2})}
\end{equation}
with the moment of inertia $\xi_{i}$ for the $i-$th fragment.

\section{III. Results and discussion}

The pre-equilibrium cluster emission in the transfer reaction is very complicated, which is not only related to the structure of the collision system e.g. the pre-formation factor, but also related to the dynamic evolution of the reaction process, i.e. the dissipation of relative motion and the coupling of internal degrees of freedom of reaction system. The emission of preequilibrium cluster is a non-equilibrium process of time and space evolution, which is a powerful probe for deeply investigating the MNT reaction dynamics.

\begin{figure*}
\includegraphics[width=16 cm]{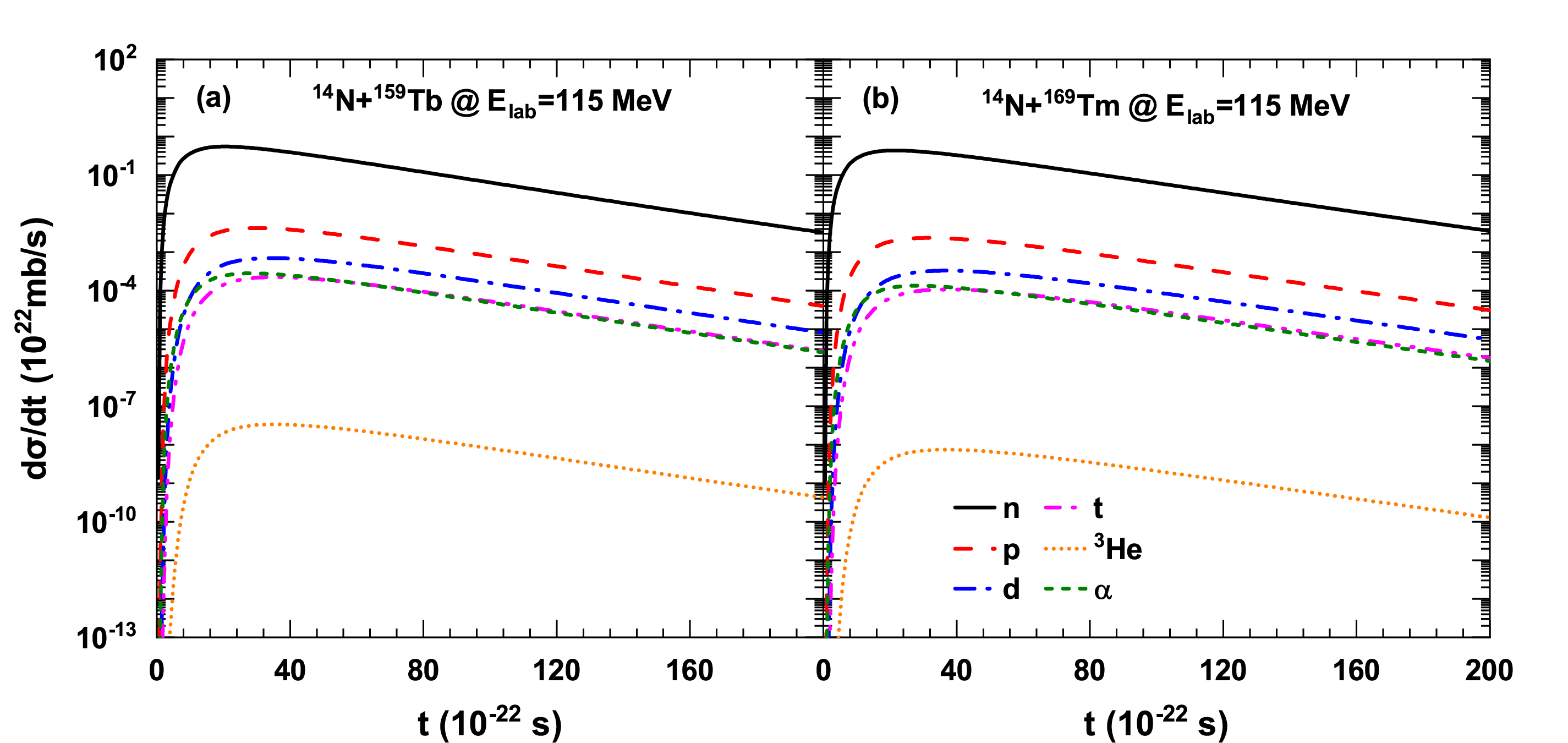}
\caption{Temporal evolution of the preequilibrium cluster emission in the reactions of (a) $^{14}$N + $^{159}$Tb and (b) $^{14}$N + $^{169}$Tm at the beam energy of 115 MeV. }
\end{figure*}

\begin{figure*}
\includegraphics[width=16 cm]{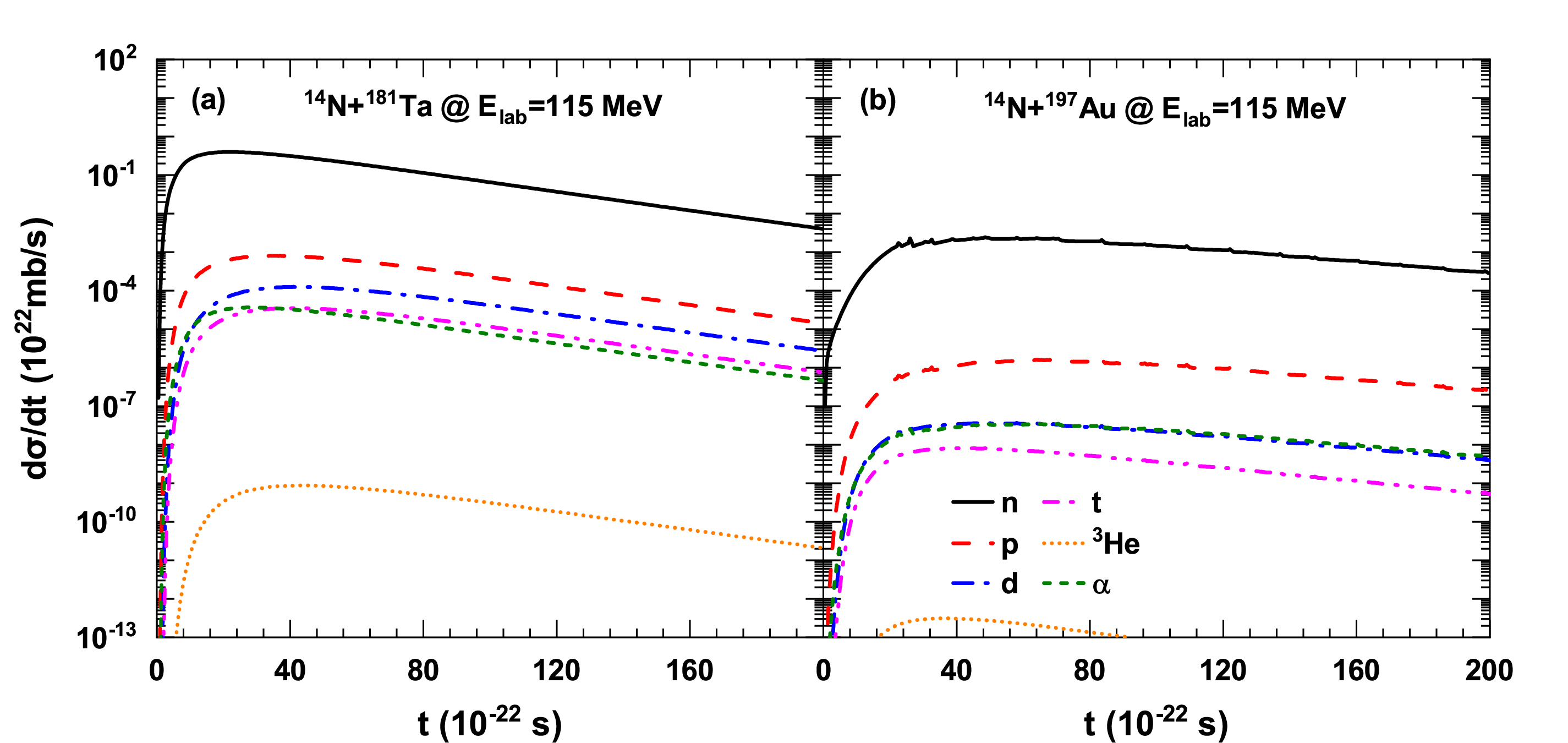}
\caption{The same as in Fig. 2, but for the collisions of $^{14}$N on $^{181}$Ta and $^{197}$Au, respectively. }
\end{figure*}

The temporal evolution of the emission probability of n, p, d, t, $^{3}$He and $\alpha$ from the transfer reactions of $^{14}$N + $^{159}$Tb, $^{169}$Tm, $^{181}$Ta and $^{197}$Au at $E_{lab}=115$ MeV is shown in Fig. 2 and Fig. 3, respectively. It is noticed that the formation of the compound nucleus is the order of a few of hundred zeptoseconds, while the reaction time of preequilibrium process is about several zeptoseconds. It can be seen from the figures that the emission of preequilibrium cluster continues until the formation of the composite nucleus. At the beginning of the reaction, the emission probability of the preequilibrium cluster increases rapidly, reaching a maximum value at about $20\sim40\times 10^{-22}$s, and then remaining stable or decreasing gradually. The emission probabilities of $\alpha$ and hydrogen isotopes are comparable, and the yields are about $3\sim4$ orders of magnitude lower than that of neutrons, but much larger than that of $^{3}$He. The local excitation energy of the DNS fragment increases with time, and the emitted clusters could take away a part of the energy, which is conducive to the formation of compound nucleus with the lower excitation energy. The total emission cross sections of the preequilibrium clusters can be obtained by counting the temporal evolution of cluster yields. The total emission cross sections of different preequilibrium particles are shown in Table\ref{tab1}. It is obvious that the $\alpha$ yields are comparable with the proton emission

\begin{table*}
\caption{\label{tab1} Production cross sections of neutron, proton, deuteron, triton, $^{3}$He and $\alpha$ in the preequilibrium process of massive transfer reactions. }
\begin{ruledtabular}
\begin{tabular}{ccccccccccccc}
&reaction system  &E$_{lab}$ (MeV)  &$\sigma_{n}$ (mb)    &$\sigma_{p}$ (mb)   &$\sigma_{d}$ (mb)  &$\sigma_{t}$ (mb)   &$\sigma_{^{3}He}$ (mb)   &$\sigma_{\alpha}$ (mb)      \\
\hline

&$^{12}$C+$^{209}$Bi   &73   &2.90$\times10^{-1}$    &2.58$\times10^{-6}$   &3.83$\times10^{-9}$   &4.06$\times10^{-8}$   &7.58$\times10^{-20}$   &4.54$\times10^{-6}$       \\

&$^{14}$N+$^{159}$Tb  &115   &1.45$\times10^{-1}$   &1.29$\times10^{-6}$   &1.92$\times10^{-9}$   &2.03$\times10^{-8}$   &3.79$\times10^{-20}$    &2.27$\times10^{-6}$    \\

&$^{14}$N+$^{169}$Tm  &115   &7.24$\times10^{-2}$    &6.46$\times10^{-7}$  &9.58$\times10^{-10}$  &1.02$\times10^{-8}$  &1.89$\times10^{-20}$    &1.13$\times10^{-6}$        \\

&$^{14}$N+$^{181}$Ta  &115   &3.62$\times10^{-2}$    &3.23$\times10^{-7}$   &4.79$\times10^{-10}$   &5.08$\times10^{-9}$   &9.74$\times10^{-21}$    &5.67$\times10^{-7}$       \\

&$^{14}$N+$^{197}$Au  &115   &1.81$\times10^{-2}$    &1.62$\times10^{-7}$   &2.40$\times10^{-10}$   &2.54$\times10^{-9}$   &4.74$\times10^{-21}$   &2.83$\times10^{-7}$        \\

&$^{14}$N+$^{209}$Bi  &115   &9.05$\times10^{-3}$    &8.08$\times10^{-8}$   &1.20$\times10^{-10}$   &1.27$\times10^{-9}$   &2.37$\times10^{-9}$   &1.42$\times10^{-7}$        \\

&$^{58}$Ni+$^{198}$Pt  &170.2   &1.11$\times10^{-4}$    &7.65$\times10^{-5}$   &7.16$\times10^{-8}$   &1.23$\times10^{-9}$   &4.30$\times10^{-10}$   &2.96$\times10^{-9}$   \\

&$^{58}$Ni+$^{198}$Pt  &185.6   &2.78$\times10^{-5}$    &1.94$\times10^{-5}$   &1.79$\times10^{-8}$   &3.08$\times10^{-10}$   &1.07$\times10^{-10}$   &7.39$\times10^{-10}$   \\

&$^{64}$Ni+$^{198}$Pt  &181.4   &6.94$\times10^{-6}$    &4.85$\times10^{-6}$   &4.47$\times10^{-9}$   &7.70$\times10^{-11}$   &2.69$\times10^{-11}$   &1.85$\times10^{-10}$   \\

&$^{72}$Ni+$^{198}$Pt  &176.0   &1.74$\times10^{-6}$    &1.21$\times10^{-6}$   &1.12$\times10^{-9}$   &1.92$\times10^{-11}$   &6.72$\times10^{-12}$   &4.62$\times10^{-11}$   \\

\end{tabular}
\end{ruledtabular}
\end{table*}

\begin{figure*}
	\includegraphics[width=16 cm]{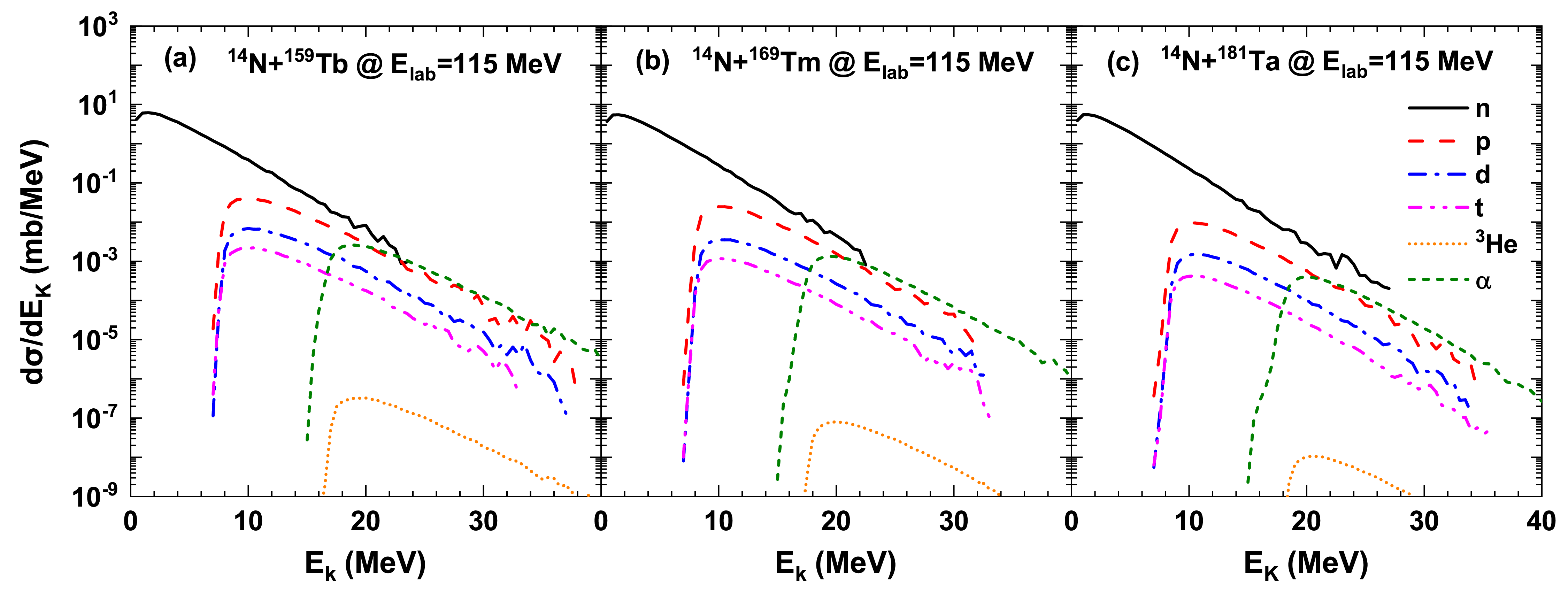}
	\caption{Kinetic energy spectra of the light nuclei produced in the reactions of (a) $^{14}$N+$^{159}$Tb, (b) $^{169}$Tm and (c) $^{181}$Ta at E$_{lab}$=115 MeV, respectively.}
\end{figure*}

Shown in Fig. 4 is the kinetic energy spectra of the light nucleus produced in the transfer reactions of $^{14}$N + $^{159}$Tb, $^{169}$Tm, $^{181}$Ta at E$_{lab}$ =115 MeV. It can be seen from the figure that the kinetic energy spectra of different reactions show the similar shape, presenting the Boltzmann distribution. The emission of neutron is the most important. Compared with the previous work \cite{Feng2023}, we have introduced the Coulomb barrier correction in this work. Hydrogen isotopes have similar emission probabilities due to the same amount of charge, and the peak of their kinetic energy spectrum is about 10 MeV. Due to $\alpha$ and $^{3}$He are more charged, the kinetic energy spectra move in the direction of greater energy (i.e. move to the right in the picture). Besides, we can see from Fig. 4 that the emission probability of $\alpha$ is about three to five orders of magnitude higher than $^{3}$He, for the reason that the former has lower separation energy and is more easy to be emitted from the DNS fragments. The above calculation results are consistent with the experimental data \cite{Ut1980, Jin1980}.

\begin{figure*}
\includegraphics[width=16 cm]{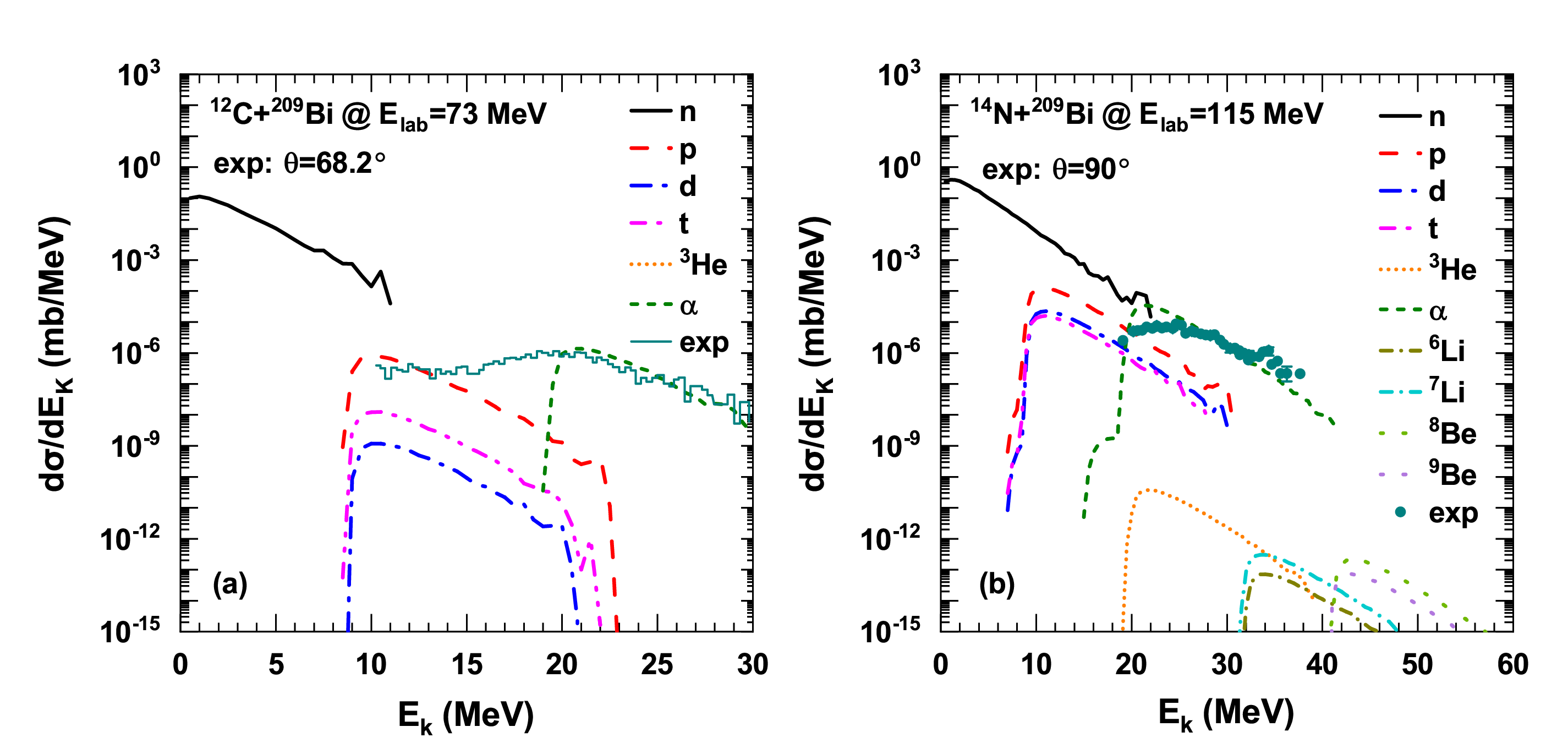}
\caption{Kinetic energy spectra of the preequilibrium clusters produced in collisions of (a) $^{12}$C+$^{209}$Bi at E$_{lab}$=73 MeV and (b) $^{14}$N+$^{209}$Bi at E$_{lab}$=115 MeV, respectively. The available experimental data are shown for comparison from HIRFL for the reaction of $^{12}$C+$^{209}$Bi \cite{Jin1980} (left panel) and from RIKEN for $^{14}$N+$^{209}$Bi \cite{Ut1980}, respectively. }
\end{figure*}

In the Fig. 5, we exhibit the kinetic energy spectra of the pre-equilibrium clusters (n, p, d, t, $^{3}$He, $\alpha$, $^{6,7}$Li, $^{8,9}$Be) in the transfer reactions induced by $^{12}$C and $^{14}$N to the same target nucleus $^{209}$Bi. The kinetic energy spectra of these pre-equilibrium particles in the transfer reactions show the nuclear structure effect and the dynamic characteristics of nuclear reaction. The available experimental data for the $\alpha$ emission from HIRFL for the massive transfer reaction of $^{12}$C+$^{209}$Bi \cite{Jin1980} and from RIKEN for $^{14}$N+$^{209}$Bi \cite{Ut1980} are nicely reproduced with the DNS model. The excitation energy of DNS fragments, and transition probability, binding energy and separation energy of transferred nucleons (clusters) affect the kinetic energy spectra. The emission cross section of preequilibrium cluster is mainly related to its formation probability and emission probability. In our calculation, it is assumed that the clusters already exist in the DNS, so the emission cross sections of different clusters are mainly determined by the emission probabilities. The higher the charge of the emitted particle, the higher the Coulomb barrier. On the other hand, the larger the separation energy of the cluster, the smaller the decay width and the lower emission probability. The kinetic energy spectra of clusters are strongly related to the Coulomb barriers and excitation energies of composite system.

\begin{figure*}
	\includegraphics[width=16 cm]{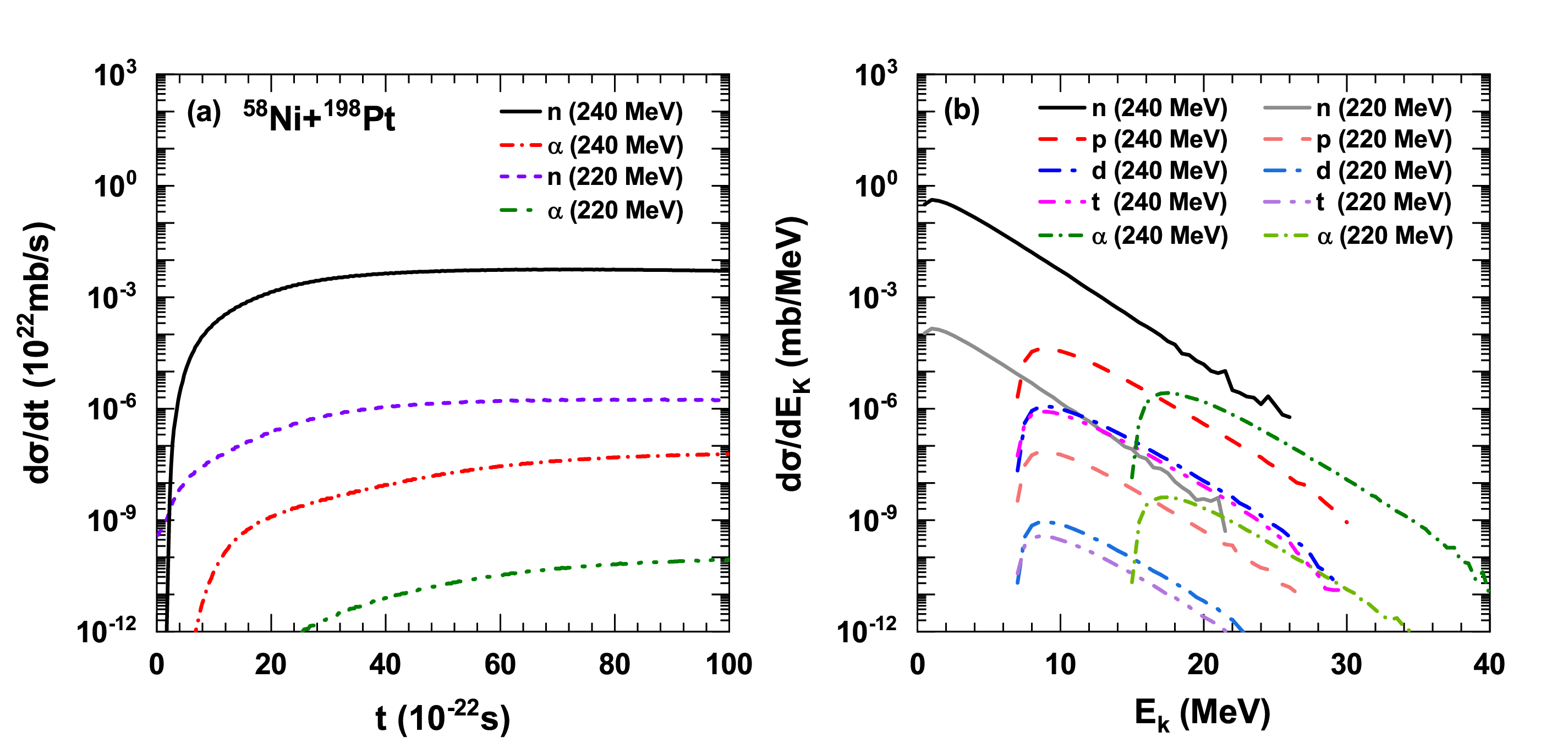}
	\caption{(a) The temporal evolution and (b) kinetic energy spectra of the preequilibrium clusters produced in collisions of $^{58}$Ni+$^{198}$Pt at E$_{c.m.}$=220 MeV and E$_{c.m.}$=240 MeV, respectively. }
\end{figure*}

\begin{figure*}
	\includegraphics[width=16 cm]{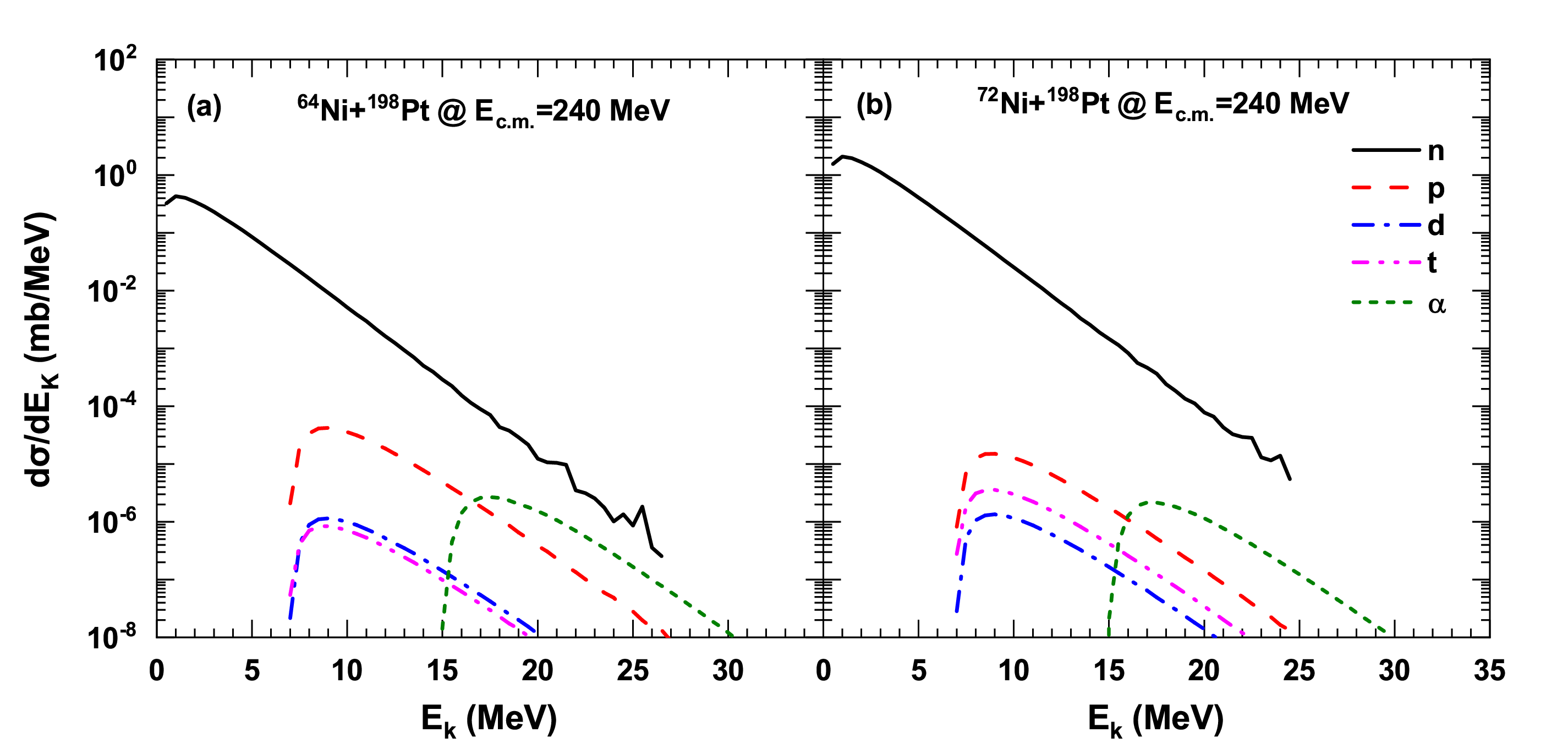}
	\caption{Kinetic energy spectra of the preequilibrium clusters produced in collisions of $^{64,72}$Ni+$^{198}$Pt at E$_{c.m.}$=240 MeV. }
\end{figure*}

The emission of the preequilibrium cluster is not only related to the reaction system, but also to the incident energy. Shown in Fig. 6 is the comparison of the time evolution and the kinetic energy distribution of the transfer reaction, $^{58}$Ni+$^{198}$Pt, at the different incident energy of 220 MeV and 240 MeV. The left part of this figure is the temporal evolution, and the right is the kinetic energy spectra of the pre-equilibrium particles. The kinetic energy of pre-equilibrium cluster is mainly determined by the local excitation energy of projectile-like and target-like fragments, and high local excitation energy is beneficial to the cluster emission. We can see that the emission probability at E$_{c.m.}$ = 240 MeV is about 2 to 3 orders of magnitude higher than at E$_{c.m.}$ = 220 MeV, indicating that the emission probability of the pre-equilibrium clusters increases with the incident energy. In Fig. 7, we compare the transfer reactions of bombarding the target nucleus $^{198}$Pt with heavier isotopes of Ni at E$_{c.m.}$ = 240 MeV, on the left is the kinetic energy spectra of the pre-equilibrium clusters emitted in the $^{64}$Ni+$^{198}$Pt reaction, and on the right is the reaction of $^{72}$Ni+$^{198}$Pt. It can be found that compared to the reaction system of $^{64}$Ni+$^{198}$Pt, the reaction induced by $^{72}$Ni seems to be more likely to emit neutrons, but the former is more likely to emit protons. In the both reaction systems, the peak of the kinetic energy spectra of proton isotopes is about 9 MeV, while the peak of $\alpha$ kinetic energy spectra is about 17 MeV.

\begin{figure*}
 \includegraphics[width=16 cm]{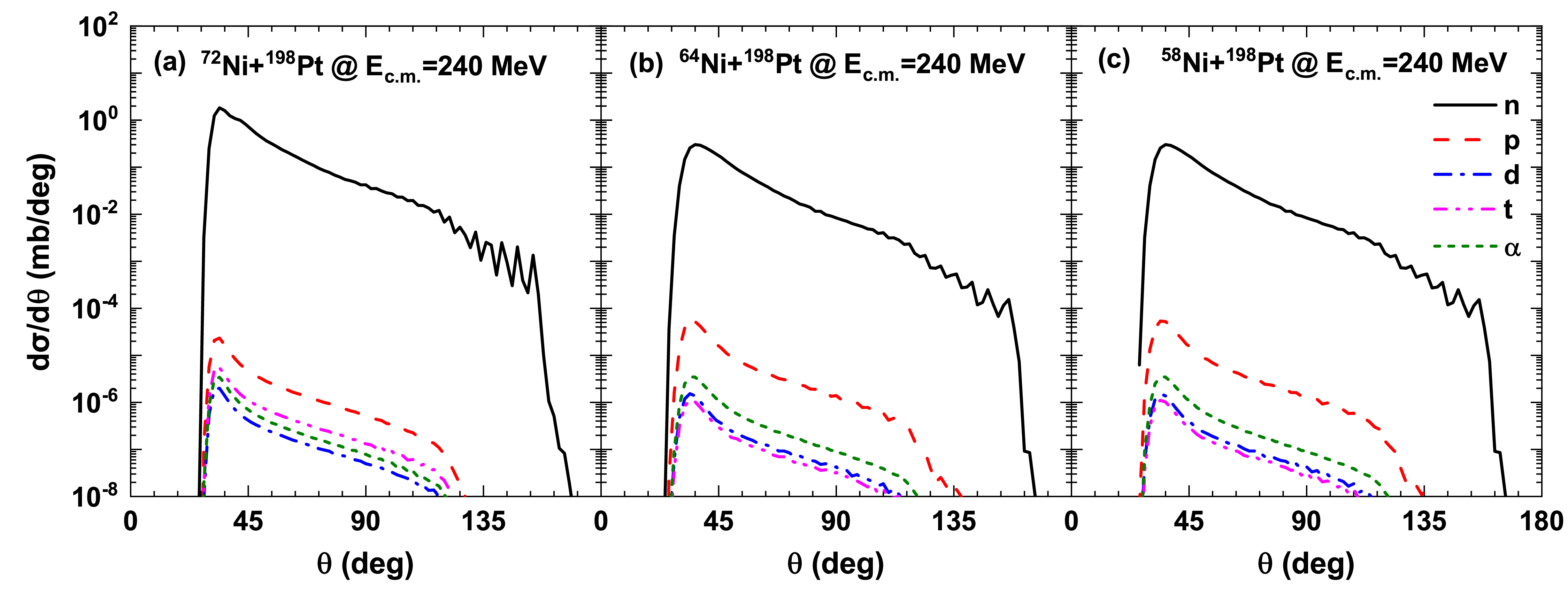}
 \caption{Angular distributions of preequilibrium clusters in the reactions of $^{58,64,72}$Ni+$^{198}$Pt at E$_{c.m.}$=240 MeV.}
\end{figure*}

The particles emitted in the preequilibrium process and from the compound nucleus have different kinetic energy and angular distributions \cite{Ho1981}. Direct particles primarily emit in the same direction as incident particles and have similar energy to each other, while the particles of the compound process emit in all directions, in equal amounts forward and backward. Preequilibrium particles tend to be emitted forward and are generally more energetic than those from the composite nucleus. Shown in Fig. 8 is the angular distributions of the emitted pre-equilibrium clusters in $^{58,64,72}$Ni+$^{198}$Pt at E$_{c.m.}$=240 MeV. The angular distribution is different for different reaction systems. For the same reaction system, the angular distributions of different clusters are roughly the same, because different clusters may be evaporated from the same excited DNS fragment. It can also be seen from the figure that the angular distributions of the pre-equilibrium particles are anisotropic, and their shape shows the similar characteristics as the angular distributions of fragments in the multinucleon transfer reactions \cite{Wo1978,Peng2022}. Under the three reaction systems, the angular distributions of the particles increase rapidly when the angle of the center of mass system is about 26$^{o}$, and there exhibits a window with 30$^{o}$-160$^{o}$ for the pre-equilibrium emission. The study of the angular distribution of pre-equilibrium clusters in the transfer reaction is of great significance to the study of the angular distribution of the primary fragments in the MNT reaction, and is helpful for the management of the experimental measurements.

\section{IV. Conclusions}
In summary, within the framework of the DNS model, we have investigated the emission mechanism of the preequilibrium clusters in the massive transfer reactions near Coulomb barrier energies, i.e., the temporal evolution, the kinetic energy spectra and the angular distributions of n, p, d, t, $^{3}$He, $\alpha$, $^{6,7}$Li, $^{8,9}$Be in collisions of $^{12}$C+$^{209}$Bi, $^{14}$N+$^{159}$Tb, $^{169}$Tm, $^{181}$Ta, $^{197}$Au, $^{209}$Bi and $^{58,64,72}$Ni+$^{198}$Pt. The cluster transfer and dynamic deformation coupled to the relative dissipation of angular momentum and motion energy in the DNS model. The emission of the preequilibrium clusters strongly depends on the incident energy, the separation energy and the Coulomb barrier from the primordial DNS fragments. The yields of hydrogen isotopes and $\alpha$ production have the similar magnitude, but more probable than the heavy particles. The kinetic energy spectra manifest the difference of charged particles, i.e., the more kinetic energy for the $\alpha$ emission than the ones of protons. The preequilibrium clusters follows the MNT fragment emission on the angular distributions and related to the correlation of nucleons. The reaction mechanism is helpful for investigating the MNT fragment formation, i.e., the yields, shell effect, emission dynamics etc, which is being planned for the forthcoming experiments at HIAF in Huizhou.

\section{Acknowledgements}
This work was supported by the National Natural Science Foundation of China (Projects No. 12175072 and No. 12311540139).

\end{document}